\documentclass[a4paper,11pt]{article}
\usepackage{latexsym}
\usepackage{feynmf}		
\def\ba{\begin{eqnarray}}
\def\ea{\end{eqnarray}}

\def\lb{\label}
\def\be{\begin{equation}}
\def\ee{\end{equation}}

\begin{document}
\baselineskip0.25in

\title{The Newton constant and gravitational waves in some vector field adjusting mechanisms}
 \author{Osvaldo P. Santill\'an \thanks{Departamento de Matem\'aticas Luis Santal\'o (IMAS), Buenos Aires, Argentina
firenzecita@hotmail.com and osantil@dm.uba.ar.} and Marina Scornavacche \thanks{Departamento de F\'isica, Universidad de Buenos Aires, Argentina
marina.scorna@hotmail.com.}}
\date {}
\maketitle

\begin{abstract}
At the present, there exist some Lorentz breaking scenarios which explain the smallness of the cosmological constant at the present era \cite{emelo0}-\cite{emelo3}. An important
aspect  to analyze is the propagation of gravitational waves and the screening or enhancement of the Newton constant $G_N$ in these models. The problem is that the Lorentz symmetry
breaking terms may induce an unacceptable value of the Newton constant $G_N$ or introduce longitudinal modes in the gravitational wave propagation. Furthermore
this breaking may spoil the standard dispersion relation $\omega=ck$. In \cite{emelo0} the authors have presented a model suggesting that the behavior of the gravitational constant is 
 correct for asymptotic times. In the present work, an explicit checking is made and we finally agree with these claims. Furthermore, it is  suggested that the gravitational waves
are also well behaved for large times. In the process, some new models with the same behavior are obtained, thus enlarging the list of possible adjustment mechanisms. 

\end{abstract}

\section{Introduction}

One of the most interesting observations of the last century is the discovery of the  cosmic acceleration. As gravity is an
attractive force, the velocity of the distant galaxies may be expected to slow down. Contrary to this,
the astronomical observations support an increasing velocity \cite{acelero1}-\cite{acelero3}. Another crucial
phenomenon is the discrepancy between the luminous matter of several objects in the universe and
their gravitational effects \cite{dm1}-\cite{dm2}. In fact, there is experimental evidence supporting a flat universe, which
implies that the actual universe energy density should be of the order of the critical one, $\rho_c\sim 10^{-47}$GeV$^4$
\cite{dm3}. This scenario does not agree with the contributions corresponding to the
dynamically measured non relativistic mass density, which is approximately $(0.1-0.3)\rho_c$. 

Several scenarios have been proposed
to explain these results. Some of them postulate the existence of dark matter. This is an unknown matter
sector whose contribution to the energy density compensates the difference between the critical and the
observed densities  \cite{dm1}-\cite{dm2}. Furthermore, the acceleration of the
universe expansion suggests the presence of a cosmological constant  \cite{acelero1}-\cite{acelero3}. If this were to be interpreted
as vacuum energy density, then its value would be a considerable fraction of the critical density $\rho_c$.

 This picture has several theoretical problems. As is well known, the predictions of different Quantum Field Theories of the vacuum energy density are at least 55 orders of magnitude larger than $\rho_c$ \cite{dolgo2}.  Thus, there exist two problems to be explained namely, why the energy density of the universe is so small and why it is so close to the critical one $\rho_c$. 

One of the approaches for solving the first problem is to assume the existence of an unknown matter component whose evolution screens the QFT energy density at late times. The first scenarios of this type include an scalar component $\varphi$ non minimally coupled to the curvature $R$ of the background space time \cite{dolgin}. The energy density $\rho(\varphi)$ of this field has a sign opposite to the critical energy $\rho_c$, in such a way that at larges times an screening of the cosmological constant takes place.
However, one of the problems that arise is that the evolution also changes the Newton constant $G_N$ to an unacceptable numerical value \cite{dolgin}. There exist a no go theorem due to Weinberg which discourages the possibility of solving
the cosmological constant problems in terms of scalar fields \cite{weinberg}. However, this theorem is formulated under certain assumptions which may be avoided in some modified gravity theories. Some examples of adjustment mechanisms in terms of scalar fields in modified gravity theories is given in \cite{urbano1}.

There exist alternative scenarios which assume that the QFT vacuum energy is not gravitating. Their underlying idea is that, if GR were the full gravity theory, then  the graviton would interact with this vacuum energy. However, if the gravity description available is only an effective one, then the gravitons may be quasi-particles and do not necessarily experience all the degrees of freedom. Quintessence models are examples of these
type of scenarios \cite{quinte}. The cosmological constant in this case is modeled by the potential energy of an unknown scalar field.

 Another type of models are the self-tuning vacuum variable scenarios \cite{volo1}. The idea behind these models is that the vacuum is a self-sustainable medium, that is, it has a definite volume even in an empty environment. These works postulate a new degree of freedom, called the q-variable, whose role is  the equilibration of the quantum vacuum. Other thermodynamical scenarios of this type were 
 considered in \cite{volo2}-\cite{volo5}. The q-scenarios inspired partially the interest in vector  vector fields adjustment mechanisms, or even tensor ones.  Some vector models capable to adjust vacuum energy to a very low value 
were presented  several years ago in \cite{dolgo1}-\cite{dolgo2}. These models spontaneously break the Lorentz symmetry, and can be considered as particular cases of more general models considered by Bjorken \cite{bjorken1}-\cite{bjorken2}. A first obstacle in the original formulation  \cite{dolgo1}-\cite{dolgo2} is that  the effective Newton constant $G_N$
obtains an unacceptable numerical value \cite{rubak}. Furthermore, they strongly modified the dispersion relation of gravitational waves and introduce longitudinal components wide beyond the experimental accuracy \cite{rubak}. However, there exist new scenarios which apparently overcome this problem \cite{emelo0}-\cite{emelo4}.

The new scenarios  \cite{emelo0}-\cite{emelo4} are precisely the motivation to write the present note. In these works it is claimed that the Newton constant $G_N$ is well behaved for large proper times $t\to\infty$. The problem is that the 
arguments these authors present are not completely accessible to our knowledge, since they require an strong knowledge of the q-models mentioned above. For this reason it is presented here an independent check about the behavior
 of the gravitational constant $G_N$ and the propagation of gravitational waves in these models. This  may be useful for a reader which is not acquainted with the ideas of \cite{volo1}-\cite{volo5}. The present work is focused in the models constructed in \cite{emelo0}, and the results presented here agree with these authors claims. The gravitational constant stabilizes to a constant value which is identified as the Newton one $G_N$. Besides, the longitudinal components of the gravitational waves vanish at $t\to\infty$.   In the process of deriving these results, general expressions for the deviation of the Planck mass $\delta M_p$ and the longitudinal components of the gravitational waves are found. Further new solutions with the same property are found as well.

The present work is organized as follows. Section 2 contains the basic equations of the model. Section 3 presents the equations for the perturbations of the model. In section 4, the Newtonian limit of these theories is analysed and an expression for the effective Planck mass is derived. Section 5 contains the equation for gravitational waves in these scenarios, and the terms that potentially may introduce longitudinal components are derived explicitly. In section 6 these features are analyzed for the models of \cite{emelo0} and it is strongly suggested that these models work fine for asymptotic times. In Section 7 new models with these properties are presented. Section 8 contains the discussion of the obtained results, and open perspectives to be investigated further are suggested.

\section{The basic equations describing the model}

The present work is related to the scenarios presented in \cite{emelo0}. The degrees of freedom of these models are two vector fields $A_\mu$ and $B_\mu$, together with the metric field $g_{\mu\nu}$. The corresponding lagrangian is given by
\be\lb{lub}
L=\epsilon(Q_A, Q_B)+L_{EH}, \qquad Q_A=\sqrt{A_{\mu;\nu}A^{\mu;\nu}}, \qquad Q_B=\sqrt{B_{\mu;\nu}B^{\mu;\nu}}.
\ee
Here $L_{EH}$ is the standard Einstein-Hilbert lagrangian and  $\epsilon(Q_A, Q_B)$ is, at this point, an arbitrary function of both scalars $Q_A$ and $Q_B$. The action is clearly non gauge invariant. The equations of motion that are obtained
by varying the action corresponding to (\ref{lub}) with respect to the vector fields $A_\mu$ and $B_\mu$ are given by
\be\lb{relax}
\nabla^\alpha(\xi_A\nabla_\alpha A_\beta)=0,\qquad \nabla^\alpha(\xi_B\nabla_\alpha B_\beta)=0.
\ee
Here the following derivatives 
$$
\xi_A=\frac{1}{2Q_A}\frac{d\epsilon}{dQ_A},\qquad\xi_B=\frac{1}{2Q_B}\frac{d\epsilon}{dQ_B},
$$
have been introduced. In the following, an isotropic and homogeneous universe will be considered, and the signature to be employed is $(+,-,-,-)$. In addition, it will be assumed
that the spatial curvature of the universe $k=0$. In this situation, there
exists a coordinate system $(t, x, y, z)$ for which the metric takes the following form
\be\lb{bang}
ds^2=dt^2-a^2(t)(dx^2+dy^2+dz^2).
\ee
Here $t$ is the proper time. The only non vanishing components of $A_\mu$ and $B_\mu$ when the space time 
metric is of the form (\ref{bang}) are the components $A_0$ and $B_0$, and
the only non vanishing Christoffel symbols for such distance element are given by 
\be\lb{relax2}
\Gamma_{ij}^0=\delta_{ij} a\dot{a},\qquad \Gamma_{0j}^i=\delta_{ij} \frac{\dot{a}}{a}.
\ee
The unique non zero derivatives of $A_0$ are given by
\be\lb{relax3}
A_{0;0}=\dot{A}_{0},\qquad A_{i;j}=- \delta_{ij}a\dot{a} A_0,\qquad 
A^{0;0}=\dot{A}_{0},\qquad A^{i;j}= -\delta_{ij}\frac{\dot{a}}{a^3} A_0.
\ee
Here the latin indices are spatial ones. Analogous formulas hold for $B_0$.

The explicit dependence of the components $A_0$ and $B_0$ with respect to the proper time $t$ is found by solving (\ref{relax}) taking into account (\ref{relax2})-(\ref{relax3}). 
The resulting equations are
\be\lb{ecuac}
\dot{\dot{A}}_0+(3H+\frac{\dot{\xi}_A}{\xi_A})\dot{A}_0-3H^2 A_0=0,
\ee
\be\lb{ecuac2}
\dot{\dot{B}}_0+(3H+\frac{\dot{\xi}_B}{\xi_B})\dot{B}_0-3H^2 B_0=0.
\ee
These equations should be supplemented with the Einstein equations. The energy momentum tensor for the vector fields  is given by
$$
T_{\alpha\beta}=\epsilon(Q_A, Q_B) g_{\alpha\beta}-2\xi_A (A_{\alpha;\gamma} A^{\;\;;\gamma}_{\beta}+A_{\gamma;\alpha} A^{\gamma}_{;\beta})-2\xi_B (B_{\alpha;\gamma} B^{\;\;;\gamma}_{\beta}+B_{\gamma;\alpha} B^{\gamma}_{;\beta})
$$
$$
+\nabla^\gamma[\xi_A(A_{\alpha} A_{\gamma;\beta}+A_{\beta} A_{\gamma;\alpha}+A_{\alpha} A_{\beta;\gamma}+A_{\beta} A_{\alpha;\gamma}-A_{\gamma} A_{\alpha;\beta}-A_{\gamma} A_{\beta;\alpha})].
$$
\be\lb{tem}
+\nabla^\gamma[\xi_B(B_{\alpha} B_{\gamma;\beta}+B_{\beta} B_{\gamma;\alpha}+B_{\alpha} B_{\beta;\gamma}+B_{\beta} B_{\alpha;\gamma}-B_{\gamma} B_{\alpha;\beta}-B_{\gamma} B_{\beta;\alpha})].
\ee
In an isotropic and homogeneous situation the only non zero components are given by
$$
\rho(A, B)=T^0_0,\qquad -P(A, B)=T^1_1=T^2_2=T^3_3,
$$
and their explicit expression can be found by use of (\ref{tem}), the result is 
$$
\rho=\epsilon(Q_A, Q_B)-Q_A \frac{d\epsilon}{dQ_A}-Q_B \frac{d\epsilon}{dQ_B},
$$
\be\lb{pree}
P=-\rho+\frac{d}{dt}\bigg(\frac{H A_0^2}{Q_A}\frac{d\epsilon}{dQ_A}\bigg)+\frac{d}{dt}\bigg(\frac{H B_0^2}{Q_B}\frac{d\epsilon}{dQ_B}\bigg)-\frac{\dot{A}_0^2}{Q_A}\frac{d\epsilon}{dQ_A}-\frac{\dot{B}_0^2}{Q_B}\frac{d\epsilon}{dQ_B}.
\ee
In finding these formulas, the equation of motions (\ref{ecuac})-(\ref{ecuac2}) should be taken into account.
Furthermore,  the following quantities
$$
Q^2_A=\dot{A}^2_0+3H^2 A_0^2,\qquad Q^2_B=\dot{B}^2_0+3H^2 B_0^2,
$$
have been introduced. In these terms the Einstein equations $R_{\alpha\beta}-g_{\alpha\beta}R=\kappa T_{\alpha \beta}$ become
$$
H^2=8\pi G_N(\Lambda+\rho(A, B)),
$$
\be\lb{in}
2\dot{H}+3H^2=8\pi G_N(\Lambda-P(A, B)).
\ee
In the following section, it will be assumed that a solution $a(t)$, $A_0(t)$ and $B_0(t)$ of (\ref{ecuac})-(\ref{ecuac2}) and (\ref{in}) has been found, and  perturbations around this vacuum will be characterized.
The presence of non zero values $A(t)$ and $B(t)$ breaks the Lorentz symmetry, since it indicates a preferred direction in the space time manifold.

\section{Generic perturbations for the model}

Consider a given solution $g_{\mu\nu}$, $A_\mu$ and $B_\mu$ of the previous model. The task in consideration is to analyze the behavior of perturbations of the form $\widetilde{g}_{\mu\nu}=g_{\mu\nu}+h_{\mu\nu}$, $\widetilde{A}_\mu=A_\mu+ c_\mu$ and $\widetilde{B}_\mu= B_\mu+d_\mu$ in the given background. Here  $h_{\mu\nu}$, $c_\mu$ and $d_\mu$ are small perturbations around the background $g_{\mu\nu}$, $A_\mu$ and $B_\mu$.
Recall that, in general, given an space time $(M, g_{\mu\nu})$ with a metric $g_{\mu\nu}$ smooth at an small neighbor of a  point $p$, the variation of the Christoffel symbols under an infinitesimal but smooth metric change $\widetilde{g}_{\mu\nu}= g_{\mu\nu}+\delta g_{\mu\nu}$ in this neighbor is given by
\be\lb{bario}
\delta \Gamma_{\mu\nu}^\alpha=\frac{g^{\alpha\kappa}}{2}(\nabla_{\mu}\delta g_{\kappa\nu}+\nabla_{\nu}\delta g_{\kappa\mu}-\nabla_{\kappa}\delta g_{\mu\nu}).
\ee
A variation of a metric $g_{\mu\nu}$ and its inverse $g^{\mu\nu}$ are not independent, in fact they are related by
$\delta g_{\mu\nu}=- g_{\mu\beta}(\delta g^{\beta\delta})g_{\delta\nu}$.  Now, given $L=\epsilon(Q_A, Q_B)$, its  expansion to second order is
$$
\delta L=\delta L_A+\delta L_B+\delta L_{AB}= \frac{1}{2}\frac{d^2\epsilon}{d Q_A^2}\delta{Q_A^2}+\frac{1}{2}\frac{d^2\epsilon}{dQ_B^2} \delta{Q_B^2}+\frac{d^2\epsilon}{d Q_B d Q_A}\delta{Q_A}\delta{Q_B}.
$$
The first term can be worked out by use of
$$
\frac{d^2\epsilon}{d Q_A^2}=Q_A\frac{d}{d Q_A}(\frac{1}{Q_A}\frac{d\epsilon}{d Q_A})+\frac{1}{Q_A}\frac{d\epsilon}{d Q_A}.
$$
In addition one has the identity
$$
\delta{Q_A}=\frac{1}{2Q_A}\delta(A_{\alpha;\beta}A^{\alpha;\beta})=\frac{1}{2Q_A}(A_{\alpha;\beta}\delta A^{\alpha;\beta}+A^{\alpha;\beta}\delta A_{\alpha;\beta})= \frac{1}{Q_A}A_{\alpha;\beta}\delta A^{\alpha;\beta}.
$$
Analogous formulas are true for $Q_B$. The variation  $\delta A_{\alpha;\beta}$ has contributions due to the fluctuations $\delta g_{\mu\nu}$ of the metric and due to the fluctuations $\delta A_\alpha$ of the vector field. Thus
\be
\delta A_{\alpha;\beta}=-\delta\Gamma_{\alpha\beta}^\gamma A_\gamma+(\delta A_{\alpha})_{;\beta}.
\ee
Here the derivative $(\delta A_{\alpha})_{;\beta}$ is taken with respect to the unperturbed metric.  In these terms
$$
\delta{Q_A^2}
=\frac{1}{Q_A^2}A^{\gamma;\delta}A^{\alpha;\beta}
[(\delta A_{\gamma})_{;\delta}(\delta A_{\alpha})_{;\beta}-2\delta\Gamma_{\gamma\delta}^\epsilon A_\epsilon (\delta A_{\alpha})_{;\beta}+\delta\Gamma_{\alpha\beta}^\epsilon \delta\Gamma_{\gamma\delta}^\eta A_\epsilon A_\eta].
$$
By taking into account the identity
$A^{\gamma;\delta}A^{\alpha;\beta}=g^{\alpha\gamma}g^{\beta\delta}Q_A^2$ it is deduced that a
$$
\delta L_A=\bigg[Q_A\frac{d}{dQ_A}(\frac{1}{Q_A}\frac{d\epsilon}{dQ_A})+\frac{1}{Q_A}\frac{d\epsilon}{dQ_A}\bigg]\frac{1}{2Q_A^2}A^{\gamma;\delta}A^{\alpha;\beta}
$$
$$
\times [(\delta A_{\gamma})_{;\delta}(\delta A_{\alpha})_{;\beta}-2\delta\Gamma_{\gamma\delta}^\epsilon A_\epsilon (\delta A_{\alpha})_{;\beta}+\delta\Gamma_{\alpha\beta}^\epsilon \delta\Gamma_{\gamma\delta}^\eta A_\epsilon A_\eta]
$$
$$
=\frac{1}{2Q_A}\bigg[\frac{d}{dQ_A}(\frac{1}{Q_A}\frac{d\epsilon}{dQ_A})A^{\gamma;\delta}A^{\alpha;\beta}+\frac{d\epsilon}{dQ_A}g^{\alpha\gamma}g^{\beta\delta}\bigg]
$$
$$
\times [(\delta A_{\gamma})_{;\delta}(\delta A_{\alpha})_{;\beta}-2\delta\Gamma_{\gamma\delta}^\epsilon A_\epsilon (\delta A_{\alpha})_{;\beta}+\delta\Gamma_{\alpha\beta}^\epsilon \delta\Gamma_{\gamma\delta}^\eta A_\epsilon A_\eta].
$$
An analogous expression follows for $\delta L_B$. The term $\delta L_{AB}$ is worked out by the identity
$$
\delta{Q_A}\delta{Q_B}=\frac{1}{Q_AQ_B}A_{\mu;\nu}
B^{\gamma;\delta}\delta A^{\mu;\nu}\delta B_{\gamma;\delta}=
\frac{1}{Q_AQ_B}A_{\mu;\nu}B^{\gamma;\delta}g^{\mu\alpha} g^{\nu\beta}\delta A_{\alpha;\beta}\delta B_{\gamma;\delta}
$$
$$
=\frac{1}{Q_AQ_B}A^{\alpha;\beta}B^{\gamma;\delta} [(\delta B_{\gamma})_{;\delta}(\delta A_{\alpha})_{;\beta}-\delta\Gamma_{\alpha\beta}^\epsilon A_\epsilon (\delta B_{\gamma})_{;\delta}-\delta\Gamma_{\gamma\delta}^\eta B_\eta \delta A_{\alpha;\beta}+\delta\Gamma_{\alpha\beta}^\epsilon \delta\Gamma_{\gamma\delta}^\eta A_\epsilon B_\eta].
$$
In these terms, it is found that
$$
\delta L_{AB}=\frac{1}{Q_AQ_B}\frac{d^2\epsilon}{d Q_B dQ_A}A^{\alpha;\beta}B^{\gamma;\delta}  [(\delta B_{\gamma})_{;\delta}(\delta A_{\alpha})_{;\beta}-\delta\Gamma_{\alpha\beta}^\epsilon A_\epsilon (\delta B_{\gamma})_{;\delta}-\delta\Gamma_{\gamma\delta}^\eta B_\eta \delta A_{\alpha;\beta}+\delta\Gamma_{\alpha\beta}^\epsilon \delta\Gamma_{\gamma\delta}^\eta A_\epsilon B_\eta].
$$
By taking into account that our background is isotropic and homogeneous, and that the only non zero vector components of the vector fields $A_\nu$ and $B_\nu$ are the time components $A_0$ and $B_0$, it follows 
that the relevant part of $\delta \Gamma^\alpha_{\beta\gamma}$ is the following
$$
\delta\Gamma_{00}^0=\frac{1}{2}h_{00,0},\qquad
\delta\Gamma_{i0}^0=\frac{1}{2}[2\frac{\dot{a}}{a}h_{i0}-h_{00,i}],
$$
\be\lb{dero}
\delta\Gamma_{ij}^0=\frac{1}{2}[-2a\dot{a}\delta_{ij} h_{00}+h_{0i,j}+h_{0j,i}-h_{ij,0}].
\ee
These expressions follow directly from (\ref{bario}) and (\ref{relax2}).  
By further making the definition
$$
\delta A_{\alpha}=c_{\alpha},\qquad
\delta B_{\alpha}=d_{\alpha},
$$
it follows that the covariant derivatives of the field perturbations are
$$
c_{0;0}=c_{0,0},\qquad c_{0;j}=c_{0,j}-\frac{\dot{a}}{a} c_j,\qquad c_{j;0}=c_{j,0}-\frac{\dot{a}}{a} c_j,\qquad c_{j;k}=c_{j,k}-\delta_{jk}a \dot{a}c_0,
$$
\be\lb{cobos}
d_{0;0}=d_{0,0},\qquad d_{0;j}=d_{0,j}-\frac{\dot{a}}{a} d_j,\qquad d_{j;0}=d_{j,0}-\frac{\dot{a}}{a} d_j,\qquad d_{j;k}=d_{j,k}-\delta_{jk}a \dot{a}d_0.
\ee
The second order variation of the lagrangian (\ref{lub}) is then given by \cite{emelo0}
$$
\delta L=\frac{1}{2Q_A}\bigg[\frac{d}{dQ_A}\bigg(\frac{1}{Q_A}\frac{d\epsilon}{dQ_A}\bigg)A^{\alpha;\beta}A^{\mu;\nu}+\frac{d\epsilon}{dQ_A}g^{\alpha\mu}g^{\beta\nu}\bigg]\bigg( c_{\alpha;\beta}c_{\mu;\nu} -A_0  c_{\alpha; \beta} \delta \Gamma^0_{\mu\nu}
+\frac{A^2_0}{4}\delta \Gamma^0_{\mu\nu}\delta \Gamma^0_{\alpha\beta}\bigg)
$$
$$
+\frac{1}{2Q_B}\bigg[\frac{d}{dQ_B}\bigg(\frac{1}{Q_B}\frac{d\epsilon}{dQ_B}\bigg)B^{\alpha;\beta}B^{\mu;\nu}+\frac{d\epsilon}{dQ_B}g^{\alpha\mu}g^{\beta\nu}\bigg]\bigg(  d_{\alpha;\beta}d_{\mu;\nu}
-B_0  d_{\alpha;\beta} \delta \Gamma^0_{\mu\nu}
+\frac{B^2_0}{4}\delta \Gamma^0_{\mu\nu}\delta \Gamma^0_{\alpha\beta}\bigg)
$$
\be\lb{lug}
+\frac{1}{Q_AQ_B}\bigg[\frac{d^2\epsilon}{dQ_A dQ_B}A^{\alpha;\beta}B^{\mu;\nu}\bigg] \bigg(c_{\alpha;\beta} d_{\mu;\nu}-\frac{B_0}{2} c_{\alpha;\beta}\delta \Gamma^0_{\mu\nu}
-\frac{A_0}{2} d_{\mu;\nu} \delta \Gamma^0_{\alpha\beta}
+\frac{A_0B_0}{4} \delta \Gamma^0_{\mu\nu}\delta \Gamma^0_{\alpha\beta}\bigg).
\ee
This second order variation should be supplemented with the Einstein-Hilbert lagrangian $\delta L_{EH}$, which is well known and its explicit form is given in any standard book of cosmology.

The importance of the perturbation lagrangian (\ref{lug}) is that it describes the propagation of inhomogeneities around the given background $g_{\mu\nu}$, $A_\mu$ and $B_\mu$. These inhomogeneities always exist, and an stern test for a realistic theory is that these
perturbations do not grow beyond the experimental accuracy. The next subsections are devoted to two important aspects of this perturbations, namely the screening or enhancement of the effective Newton constant $G_N$ and the propagation of non transversal modes for gravitational waves. These effects should be suppressed for these theories, otherwise these model would not be realistic.

\section{The screening or enhancement of the Planck mass $M_p$}
In the following, the Newton limit of the general class of theories described above will be derived. In other words, it will be assumed that there is a spherically symmetric matter source, such an spherical planet.
In the non relativistic limit, the gravitational field is described by the Poisson equation
\be\lb{pois}
M_p^2\Delta \Phi=T_{00}.
\ee
Here $\Phi$ is the Newton potential of standard classical mechanics and the Newton constant is identified as $G_N\sim M_p^{-2}$
in natural units. The energy momentum tensor $T_{00}$ is due to the source, and $T_{00}\sim \delta(x)\delta(y)\delta(z)$, that is, the source is a mass point located at the origin. 

Now, in the cosmological setup we are working with, the solutions $A_0(t)$, $B_0(t)$ and $a(t)$ of the system (\ref{ecuac})-(\ref{ecuac2}) and (\ref{in}) are functions of the proper time $t$.
However, in the following, it will be assumed that this time dependence is not relevant at the present times and these quantities may be considered as almost constants. The idea is to include an almost static mass source in this static background, and to calculate the components $h_{\mu\nu}$ due to this perturbation.
It is important to remark that in GR, the Newtonian potential $\Phi$ is identified as usual by the relation
$g_{00}=1+2\Phi$ from where it follows that
\be\lb{ido}
\frac{2\Phi}{c}=h_{00}.
\ee
Thus, special attention should be paid to the component $h_{00}$. The time dependence of these quantities will be assumed to be very slow. Now, the terms in the lagrangian (\ref{lug}) which contains the perturbations $c_\alpha$ are generically of the form
$$
\delta L_c=r \delta^{ag}\delta^{bd}(c_{a;b}c_{g;d}-A_0c_{a;b}(h_{0g;d}+h_{0d;g}-h_{gd;0}))
+p \delta^{ab}\delta^{gd}(c_{a;b}c_{g;d}-A_0c_{a;b}(h_{0g;d}+h_{0d;g}-h_{gd;0}))
$$
\be\lb{pirtu}
+s \delta^{bd}(c_{0;b}c_{0;d}-A_0c_{0;b}h_{00;d})+q \delta^{ab}\delta^{gd}(c_{a;b}d_{g;d}-\frac{B_0}{2}c_{a;b}(h_{0g;d}+h_{0d;g}-h_{gd;0})).
\ee
By taking (\ref{relax3}) into account, it is found the following expression for the coefficients $p$, $q$, $r$ and $s$ 
$$
p=\frac{1}{2Q_A}\frac{d}{dQ_A}\bigg(\frac{1}{Q_A}\frac{d\epsilon}{dQ_A}\bigg)\frac{H}{a^2}A_0^2,
$$
\be\lb{cofe}
q=\frac{1}{Q_A Q_B}\frac{d^2\epsilon}{dQ_A dQ_B}\frac{H}{a^2}A_0 B_0,\qquad r=\frac{1}{2Q_A}\frac{d\epsilon}{dQ_A},\qquad s=-2r.
\ee
All these functions are evaluated at the present time $t_0$. As stated above, these coefficients will be considered as simple constants, this assumption will be justified later on. Here the latin indices
are spatial ones and greek indices  are generic, that is, they may denote an spatial or time component. For the Newtonian limit, one may choose the gauge $h_{0g;}^{\;\;\;\;\;g}=0$. In fact, one have that
$$
h_{0g;}^{\;\;\;\;g}=\frac{1}{a^2}\partial_g h_{0g}-3H h_{00}-\frac{3H}{a^2}h_{gg}.
$$
From this expression, by taking into account that $H\to 0$ for $t\to \infty$ it follows that this gauge 
is approximately $h_{0g,g}=0$, which is the Coulomb gauge usually employed for studying perturbations in Minkowski space time. Furthermore, since the source is an static mass, the time dependence of the components $h_{\alpha\beta}$, $c_\alpha$ and $d_\alpha$ will be neglected. In particular, from the expressions
\be\lb{expres}
h_{00;0}=h_{00,0},\qquad h_{0i;0}=h_{0i,0}+H h_{0i},\qquad h_{ij;0}=h_{ij,0}+2H h_{ij},
\ee
and by taking into account that  $H\to 0$ for large times, it follows that the covariant derivatives $h_{\alpha\beta;0}$ may be neglected as well. The same follows for the covariant derivatives $c_{\alpha;0}$, as follows from (\ref{cobos}). These are all the assumptions to be used when deriving the equations of motions
$$
\nabla_\beta \bigg(\frac{\delta L}{\delta\nabla_\beta c_\alpha}\bigg)=0,\qquad \nabla_\beta \bigg(\frac{\delta L}{\delta\nabla_\beta d_\alpha}\bigg)=0,\qquad \nabla_\beta \bigg(\frac{\delta L}{\delta\nabla_\beta h_{\alpha\gamma}}\bigg)=0
$$

The equation of motion for the component $c_0$ is given by
\be\lb{espin1}
2 c_{0;ii}-A_0 h_{00;ii}=0.
\ee
The analogous equation follows for $d_0$. Thus, one has that
$$
2c_0-A_0 h_{00}=f,
$$
with $f$ such that $\Delta f=0$. This harmonic function should be zero at infinite.
By further assuming that when $h_{00}\to0$ then $c_0\to 0$ and $d_0\to 0$, it is reasonable to assume that $f$ should be zero everywhere
and thus
\be\lb{csr2}
c_{0}=\frac{A_0}{2}h_{00},\qquad d_{0}=\frac{B_0}{2}h_{00}.
\ee
In addition, the corresponding equations for the spatial components $c_i$ are given by
$$
q \delta^{ab}\delta^{gd}(\delta_{ai}\delta_{bj}d_{g;dj}-\frac{B_0}{2}\delta_{ai}\delta_{bj}(h_{0g;dj}+h_{0d;gj}))
$$
$$
+p \delta^{ab}\delta^{gd}(\delta_{ai}\delta_{bj} c_{g;dj}+c_{a;bj}\delta_{gi}\delta_{dj} -A_0\delta_{ai}\delta_{bj}(h_{0g;dj}+h_{0d;gj}))
$$
$$
+r \delta^{ag}\delta^{bd}(\delta_{ai}\delta_{bj} c_{g;dj}+c_{a;bj}\delta_{gi}\delta_{dj} -A_0\delta_{ai}\delta_{bj}(h_{0g;dj}+h_{0d;gj}))=0,
$$
which, after simplifying the Kronecker deltas, becomes
\be\lb{espin}
p(2 c_{g;gi}-2A_0h_{0g;gi})+q(d_{g;gi}-B_0 h_{0g;gi})+r(2c_{i;gg}-A_0 (h_{0i;gg}+h_{0g;ig}))=0.
\end{equation}
Analogous equations hold for the component $d_{i}$. By taking into account the gauge $h_{0g;}^{\;\;\;\;g}=0$, these equations are solved by
\be\lb{csr}
c_{i}=\frac{A_0}{2} h_{0i},\qquad d_{i}=\frac{B_0}{2} h_{0i}.
\ee
This can be seen as follows. The condition $h_{0g;}^{\;\;\;\;g}=0$ together with (\ref{csr}) imply that $c_{g;}^{\;\;\;g}=d_{g;}^{\;\;\;g}=0$. If this conclusion is introduced into (\ref{espin}) the resulting equation is
$$
c_{i;gg}-\frac{A_0}{2} h_{0i;gg}=0.
$$
It is direct to check  that (\ref{csr}) is solution of this equation, which is what we wanted to show. The same argument applies for the $d_i$ components.

The final task is to consider the part of the lagrangian (\ref{lug}) corresponding to the graviton  $h_{ij}$. This is explicitly given by
$$
\delta L_h=p \delta^{ab}\delta^{gd}(-A_0c_{a;b}(h_{0g;d}+h_{0d;g}-h_{dg;0})+\frac{A_0^2}{4}(h_{0a;b}+h_{0b;a}-h_{ab;0})(h_{0g;d}+h_{0d;g}-h_{dg;0}))
$$
$$
+t \delta^{ab}\delta^{gd}(-B_0d_{a;b}(h_{0g;d}+h_{0d;g}-h_{dg;0})+\frac{B_0^2}{4}(h_{0a;b}+h_{0b;a}-h_{ab;0})(h_{0g;d}+h_{0d;g}-h_{dg;0}))
$$
$$
-q \delta^{ab}\delta^{gd}(\frac{B_0}{2}c_{a;b}(h_{0g;d}+h_{0d;g}+h_{dg;0})-\frac{A_0}{2}d_{g;d}(h_{0a;b}+h_{0b;a}-h_{ab;0})
$$
$$
-\frac{A_0B_0}{4}(h_{0a;b}+h_{0b;a}-h_{ab;0})(h_{0g;d}+h_{0d;g}-h_{dg;0}))
$$
$$
+r \delta^{ag}\delta^{bd}(-A_0c_{a;b}(h_{0g;d}+h_{0d;g}-h_{dg;0})+\frac{A_0^2}{4}(h_{0a;b}+h_{0b;a}-h_{ab;0})(h_{0g;d}+h_{0d;g}-h_{dg;0}))
$$
$$
+u \delta^{ag}\delta^{bd}(-B_0d_{a;b}(h_{0g;d}+h_{0d;g}-h_{dg;0})+\frac{B_0^2}{4}(h_{0a;b}+h_{0b;a}-h_{ab;0})(h_{0g;d}+h_{0d;g}-h_{dg;0}))
$$
\be\lb{lug2}
+s \delta^{bd}(-A_0c_{0;b}h_{00;d}+\frac{A^2_0}{4}h_{00;b}h_{00;d})+v \delta^{bd}(-B_0d_{0;b}h_{00;d}+\frac{B^2_0}{4}h_{00;b}h_{00;d})+\delta L_{EH}.
\ee
Here the coefficients $p$, $q$, $r$  and $s$ are given in (\ref{cofe}) and the remaining ones follows from (\ref{lug}) and (\ref{relax3}). Their explicit expression is
\be\lb{cofe2}
u=\frac{1}{2Q_B}\frac{d\epsilon}{dQ_B},\qquad
v=-2u,\qquad
 t=\frac{1}{2Q_B}\frac{d}{dQ_B}\bigg(\frac{1}{Q_B}\frac{d\epsilon}{dQ_B}\bigg)\frac{H}{a^2}B_0^2,
\ee
Note that the deltas $\delta^{ij}$ are related to spatial indices, therefore the only contributions to the equations of motion comes from
the terms proportional to $s$ and $v$. By taking into account (\ref{cofe}), the identification (\ref{csr}) and the gauge $h_{0,gg}=0$ the resulting equations corresponding to $h_{00}$
are given by
\be\lb{ido2}
\bigg(\frac{M_p^2}{2}+\frac{A^2_0}{2Q_A}\frac{d\epsilon}{dQ_A}+\frac{B^2_0}{2Q_B}\frac{d\epsilon}{dQ_B}\bigg)
\nabla^i\nabla_i h_{00}=T_{00}.
\ee
Here 
$$
\nabla^i\nabla_i h_{00}=\frac{\delta^{ij}}{a^2}\partial_i \partial_jh_{00}+\Gamma^0_{ii}\dot{h}_{00}=\frac{\delta^{ij}}{a^2}\partial_i \partial_jh_{00},
$$
the last identity follows from the fact that the time dependence of the metric $h_{00}$ can be neglected and also that $\Gamma^0_{ij}\sim H\to 0$ for $t\to \infty$.
By taking into account that $2\Phi=h_{00}$, as discussed above in (\ref{ido}),  the last equation (\ref{ido2}) resembles the Poisson equation (\ref{ido})
with an effective Planck mass given by
\be\lb{effectmass}
M_{eff}^2=M_p^2+2\delta M_p^2=M_p^2+\frac{A^2_0}{Q_A}\frac{d\epsilon}{dQ_A}+\frac{B^2_0}{Q_B}\frac{d\epsilon}{dQ_B}.
\ee
Thus, one of the conditions for these models to be realistic is that the quantity 
\be\lb{real}
\delta M_p^2=\frac{A^2_0}{Q_A}\frac{d\epsilon}{dQ_A}+\frac{B^2_0}{Q_B}\frac{d\epsilon}{dQ_B},
\ee
goes to zero or to a constant for large times.

The other point to be checked is that  $h_{0i}=0$ and $h_{ij}=0$ are solutions of the model, otherwise an spherically mass may induce anisotropies.  The equations for these degrees of freedom derived from (\ref{lug2}) and are identically zero due to the gauge condition $h_{0,i}^i=0$.
Thus this type of spherically symmetrical solutions are allowed in the model.

\section{Gravitational waves}

The next important aspect to study is the propagation of gravitational wave in the given Lorentz breaking background. In this case, the components
 $h_{\mu\nu}$, $c_\mu$ and $d_\mu$ are assumed to be time dependent. By taking into account (\ref{expres}), it is seen that $h_{\alpha\beta;0}=h_{\alpha\beta,0}$ for large times. But this will be imposed at the end of the calculations. The lagrangian (\ref{lug}) in these circumstances is given by
$$
L=\beta\bigg[c_{i;i}c_{j;j}-A_0 c_{i;i}(2h_{0j;j}-h_{jj;0})+\frac{A_0^2}{4}(2h_{0i;i}-h_{ii;0})(2h_{0j;j}-h_{jj;0})\bigg]
$$
$$
+b\bigg[d_{i;i}d_{j;j}-B_0 d_{i;i}(2h_{0j;j}-h_{jj;0})+\frac{B_0^2}{4}(2h_{0i;i}-h_{ii;0})(2h_{0j;j}-h_{jj;0})\bigg]
$$
$$
+\epsilon\bigg[c_{i;i}c_{0;0}-A_0 c_{i;i}h_{00;0}+\frac{A_0^2}{4}(2h_{0i;i}-h_{ii;0})h_{00;0}\bigg]
$$
$$
+\delta\bigg[c_{i;i}c_{0;0}-A_0 c_{0;0}(2h_{0i;i}-h_{ii;0})+\frac{A_0^2}{4}(2h_{0i;i}-h_{ii;0})h_{00;0}\bigg]
$$
$$
+e\bigg[d_{i;i}d_{0;0}-B_0 d_{i;i}h_{00;0}+\frac{B_0^2}{4}(2h_{0i;i}-h_{ii;0})h_{00;0}\bigg]
$$
$$
+d\bigg[d_{i;i}d_{0;0}-B_0 d_{0;0}(2h_{0i;i}-h_{ii;0})+\frac{B_0^2}{4}(2h_{0i;i}-h_{ii;0})h_{00;0}\bigg]
$$
$$
+\xi\bigg[c_{i;i}d_{0;0}-\frac{B_0}{2} c_{i;i}h_{00;0}-\frac{A_0}{2} d_{i;i}h_{00;0}+\frac{A_0B_0}{4}(2h_{0i;i}-h_{ii;0})h_{00;0}\bigg]
$$
$$
+\kappa\bigg[c_{0;0}d_{i;i}-\frac{B_0}{2} c_{0;0}(2h_{0i;i}-h_{ii;0})-\frac{A_0}{2} d_{0;0}(2h_{0i;i}-h_{ii;0})+\frac{A_0B_0}{4}(2h_{0i;i}-h_{ii;0})h_{00;0}\bigg]
$$
\be\lb{bush}
+\lambda\bigg[c_{0;0}d_{0;0}-\frac{B_0}{2} c_{0;0}h_{00;0}-\frac{A_0}{2} d_{0;0}h_{00;0}+\frac{A_0B_0}{4}h_{00;0}h_{00;0}\bigg]+\delta L_{EH}
\ee
$$
+\sigma\bigg[c_{0;0}c_{0;0}-A_0 c_{0;0}h_{00;0}+\frac{A_0^2}{4}h_{00;0}h_{00;0}\bigg]+s\bigg[d_{0;0}d_{0;0}-B_0 d_{0;0}h_{00;0}+\frac{B_0^2}{4}h_{00;0}h_{00;0}\bigg]
$$
$$
+\gamma\bigg[c_{i;i}d_{j;j}-\frac{B_0}{2} c_{i;i}(2h_{0j;j}-h_{jj;0})-\frac{A_0}{2} d_{i;i}(2h_{0i;i}-h_{ii;0})+\frac{A_0 B_0}{4}(2h_{0i;i}-h_{ii;0})(2h_{0j;j}-h_{jj;0})\bigg]
$$
$$
+\alpha\bigg[c_{\mu;\nu}c^{\mu; \nu}-A_0 c^{\mu;\nu}(h_{0\mu; \nu}+h_{0\nu; \mu}-h_{\mu\nu;0})+\frac{A_0^2}{4}(h_{0\mu; \nu}+h_{0\nu; \mu}-h_{\mu\nu;0})(h^{0\mu; \nu}+h^{0\nu; \mu}-h^{\mu\nu;0})\bigg]
$$
$$
+a\bigg[d_{\mu;\nu}d^{\mu; \nu}-B_0 d^{\mu;\nu}(h_{0\mu; \nu}+h_{0\nu; \mu}-h_{\mu\nu;0})+\frac{B_0^2}{4}(h_{0\mu; \nu}+h_{0\nu; \mu}-h_{\mu\nu;0})(h^{0\mu; \nu}+h^{0\nu; \mu}-h^{\mu\nu;0})\bigg]
$$
The coefficients in the previous expression can be obtained by comparing (\ref{relax3}) and (\ref{lug}). The explicit  result is given by
$$
\alpha=\frac{1}{2Q_A}\frac{d\epsilon}{dQ_A},\qquad a=\frac{1}{2Q_B}\frac{d\epsilon}{dQ_B},\qquad \beta=\frac{1}{2Q_A}\frac{d}{dQ_A}\bigg(\frac{1}{Q_A}\frac{d\epsilon}{dQ_A}\bigg)\frac{\dot{a}^2}{a^6}A_0^2,
$$
$$
b=\frac{1}{2Q_B}\frac{d}{dQ_B}\bigg(\frac{1}{Q_B}\frac{d\epsilon}{dQ_B}\bigg)\frac{\dot{a}^2}{a^6}B_0^2,\qquad
\epsilon=\delta=-\frac{1}{2Q_A}\frac{d}{dQ_A}\bigg(\frac{1}{Q_A}\frac{d\epsilon}{dQ_A}\bigg)\frac{\dot{a}}{a^3}A_0 \dot{A}_0,
$$ 
$$
e=d=-\frac{1}{2Q_B}\frac{d}{dQ_B}\bigg(\frac{1}{Q_B}\frac{d\epsilon}{dQ_B}\bigg)\frac{\dot{a}}{a^3}B_0 \dot{B}_0,\qquad
\sigma=\frac{1}{2Q_A}\frac{d}{dQ_A}\bigg(\frac{1}{Q_A}\frac{d\epsilon}{dQ_A}\bigg)\dot{A}_0^2,
$$
$$
s=\frac{1}{2Q_B}\frac{d}{dQ_B}\bigg(\frac{1}{Q_B}\frac{d\epsilon}{dQ_B}\bigg)\dot{B}_0^2,\qquad \xi=-\frac{1}{Q_AQ_B}\frac{d^2\epsilon}{dQ_AdQ_B}\frac{\dot{a}}{a^3} A_0 \dot{B}_0,
$$
\be\lb{busho}
 \kappa=-\frac{1}{Q_AQ_B}\frac{d^2\epsilon}{dQ_AdQ_B}\frac{\dot{a}}{a^3} \dot{A}_0 B_0,\qquad
 \lambda=\frac{1}{Q_AQ_B}\frac{d^2\epsilon}{dQ_AdQ_B} \dot{A}_0 \dot{B}_0.
\ee
A convenient gauge for analyzing gravitational waves propagation is  $d_0=h_{i0}=0$. The first condition can be fixed by use of convenient Lorentz transformation while the second is obtained by use of spatial rotations. This gauge was already considered in \cite{rubak} when studying this type of models. By taking this gauge into account the equations of motion for $c_0$ and $d_0$ may be expressed as follows
$$
[2\sigma c_{0;0}+(\epsilon+\delta)c_{i;i}+\kappa d_{i;i}]_{;0}=-\bigg[\bigg(\delta A_0+\frac{\kappa B_0}{2}\bigg)h_{ii;0}-\bigg(\sigma A_0+\frac{\lambda B_0}{2}\bigg)h_{00;0}\bigg]_{;0}-\bigg(\frac{\alpha A_0}{2}h_{00;\alpha}\bigg)^{;\alpha},
$$
\be\lb{inoa1}
[\lambda c_{0;0}+(e+d)d_{i;i}+\kappa c_{i;i}]_{;0}=-\bigg[\bigg(d B_0+\frac{\kappa A_0}{2}\bigg)h_{ii;0}-\bigg(s B_0+\frac{\lambda A_0}{2}\bigg)h_{00;0}\bigg]_{;0}-\bigg(\frac{a B_0}{2}h_{00;\alpha}\bigg)^{\alpha}.
\ee
By use of the same gauge, the equations for the spatial components $c_i$ and $d_i$ are respectively given by
$$
[(\epsilon+\delta)c_{0;0}]_{;0}+2 (\alpha c_{i;\beta})^{;\beta}-\bigg(\frac{\alpha A_0}{2}h_{00;i}\bigg)_{;0}+\frac{\alpha A_0}{2}h_{ij;0j}
$$
$$
+\beta\bigg(2 c_{j;ji}+A_0h_{jj;0i}\bigg)
+\gamma\bigg(d_{j;ji}+\frac{B_0}{2}h_{jj;0j}\bigg)-(\epsilon A_0 +\xi \frac{B_0}{2})h_{00;0i}=0,
$$
$$
(\kappa c_{0;0})_{;0}+2 (a d_{i;\beta})^{;\beta}-\bigg(\frac{a B_0}{2}h_{00;i}\bigg)_{;0}+\frac{a B_0}{2}h_{ij;0j}
$$
\be\lb{inoa2}
+b\bigg(2 d_{j;ji}+B_0h_{jj;0i}\bigg)
+\gamma\bigg(c_{j;ji}+\frac{A_0}{2}h_{jj;0i}\bigg)-(e B_0+\xi \frac{A_0}{2})h_{00;0i}=0.
\ee
The equation for $h_{00}$ is
$$
-(\alpha A_0 c^{(0;\alpha)})_{;\alpha}
+(\alpha A_0^2 h^{00;\alpha})_{;\alpha}-(\frac{\alpha A_0^2}{2}h^{00;0})_{;0}
-(a B_0 d^{(0;\alpha)})_{;\alpha}
+(a B_0^2 h^{00;\alpha})_{;\alpha}-(\frac{a B_0^2}{2}h^{00;0})_{;0}
$$
$$
-(\epsilon A_0 c_{i;i})_{;0}-\bigg(\frac{\epsilon A_0^2}{4}h_{ii;0}\bigg)_{;0}-\bigg(\frac{\delta A_0^2}{4}h_{ii;0}\bigg)_{;0}
-(eB_0 d_{i;i})_{;0}-\bigg(\frac{e B_0^2}{4}h_{ii;0}\bigg)_{;0}-\bigg(\frac{d B_0^2}{4}h_{ii;0}\bigg)_{;0}
$$
$$
+\bigg(\frac{\sigma A_0^2}{2}h_{00;0}\bigg)_{;0}+\bigg(\frac{s B_0^2}{2}h_{00;0}\bigg)_{;0}
-\bigg(\frac{\xi B_0}{2} c_{i;i}\bigg)_{;0}-\bigg(\frac{\xi A_0}{2} d_{i;i}\bigg)_{;0}-\bigg(\frac{\xi A_0B_0}{4}h_{ii;0}\bigg)_{;0}
$$
\be\lb{inoa3}
-(\sigma A_0 c_{0;0})_{;0}-(\lambda B_0 c_{0;0})_{;0}-\bigg(\frac{\kappa A_0B_0}{4}h_{ii;0}\bigg)_{;0}
+\bigg(\frac{\lambda A_0B_0}{2}h_{00;0}\bigg)_{;0}+E_{00}=0.
\ee
Here the term $E_{00}$ is the contribution of $\delta L_{EH}$. For the component $h_{0k}$ one has that
$$
- (\alpha A_0c^{(k;\alpha)})_{;\alpha}+(\alpha A_0c^{(0;\alpha)})_{;0}
+\bigg(\frac{\alpha A_0^2}{2}h^{0k; \alpha}\bigg)_{;\alpha}+\bigg(\frac{\alpha A_0^2}{2}h^{0\alpha;k}\bigg)_{;\alpha}-\bigg(\frac{\alpha A_0^2}{2}h^{k\alpha;0}\bigg)_{;\alpha}-\bigg(\frac{\alpha A_0^2}{2}h^{00;k}\bigg)_{;0}
$$
$$
- (aB_0d^{(k;\alpha)})_{;\alpha}+(a B_0d^{(0;\alpha)})_{;0}
+\bigg(\frac{a B_0^2}{2}h^{0k; \alpha}\bigg)_{;\alpha}+\bigg(\frac{a B_0^2}{2}h^{0\alpha;k}\bigg)_{;\alpha}-\bigg(\frac{a B_0^2}{2}h^{k\alpha;0}\bigg)_{;\alpha}-\bigg(\frac{a B_0^2}{2}h^{00;k}\bigg)_{;0}$$
$$
+\frac{eB_0^2}{4}h_{00;0k}+\frac{dB_0^2}{4}h_{00;0k}
-\gamma\bigg(\frac{B_0}{2} c_{i;ik}+\frac{A_0}{2} d_{i;ik}+\frac{A_0 B_0}{2}h_{jj;0k}\bigg)
+\frac{\xi A_0B_0}{4}h_{00;0k}
$$
\be\lb{inoa4}
+\frac{\kappa A_0B_0}{4}h_{00;0k}-\beta\bigg(A_0 c_{i;ik}+\frac{A_0^2}{2}h_{jj;0k}\bigg)
-b\bigg(B_0 d_{i;ik}+\frac{B_0^2}{2}h_{jj;0k}\bigg)
\ee
$$
-(2\delta A_0+\kappa B_0) c_{0;0k}+\frac{\epsilon A_0^2}{4}h_{00;0k}+\frac{\delta A_0^2}{4}h_{00;0k}+E_{0k}=0.
$$
Finally, for $h_{ik}$, it is obtained that
$$
(\alpha A_0c^{(k;l)})_{;0}+\bigg(\frac{\alpha A_0^2}{2}h^{lk;0}\bigg)_{;0}
+(aB_0d^{(k;l)})_{;0}+\bigg(\frac{a B_0^2}{2}h^{lk;0}\bigg)_{;0}
$$
$$
+(\beta A_0 c_{i;i})_{;0}\delta_{kl}+\bigg(\frac{\beta A_0^2}{2}h_{jj;0}\bigg)_{;0}\delta_{kl}
+(b B_0 d_{i;i})_{;0}\delta_{kl}+\bigg(\frac{b B_0^2}{2}h_{jj;0}\bigg)_{;0}\delta_{kl}-\bigg(\frac{\epsilon A_0^2}{4}h_{00;0}\bigg)_{;0}
\delta_{kl}$$
$$
-\bigg(\frac{eB_0^2}{4}h_{00;0}\bigg)_{;0}\delta_{kl}-\bigg(\frac{dB_0^2}{4}h_{00;0}\bigg)_{;0}\delta_{kl}
+\bigg(\frac{\gamma B_0}{2} c_{i;i}\bigg)_{;0}\delta_{kl}+\bigg(\frac{\gamma A_0}{2} d_{i;i}\bigg)_{;0}\delta_{kl}+\bigg(\frac{\gamma A_0 B_0}{2}h_{jj;0}\bigg)_{;0}\delta_{kl}
$$
\be\lb{inoa5}
-[(2\delta A_0+\kappa B_0) c_{0;0}]_{;0}\delta_{kl}+\bigg(\frac{\delta A_0^2}{4}h_{00;0}\bigg)_{;0}\delta_{kl}-\bigg(\frac{\xi A_0B_0}{4}h_{00;0}\bigg)_{;0}\delta_{kl}
-\bigg(\frac{\kappa A_0B_0}{4}h_{00;0}\bigg)_{;0}\delta_{kl}+E_{kl}=0.
\ee
As stated above, the quantities $E_{\alpha\beta}$  in equations (\ref{inoa3})-(\ref{inoa5}) are the contributions for the standard perturbed Einstein-Hilbert lagrangian $\delta L_{EH}$.
If this terms were only present, then the solution of the model will be the propagation of gravitational waves, which are transversal and has the dispersion relation $\omega=k$.
It is the presence of the additional terms the ones which may introduce longitudinal modes or to modify the dispersion relation. Thus, another stern requirement for a realistic
theory is that the effect of these additional terms is negligible for large times $t\to \infty$.

\section{Tests for known solutions}
Having derived the deviation from the mass Planck mass $\delta M_p$ in (\ref{real}) and the deviations from the gravitational wave equations in (\ref{inoa3})-(\ref{inoa5}),
the next step is to test the known solutions of the model.
In the work \cite{emelo0} the function $\epsilon(Q_A, Q_B)$ the authors consider is given by
\be\lb{alpe}
\epsilon=\frac{Q_A^4-Q_B^4}{Q_A^2 Q_B^2+\delta M_p^8},
\ee
with $\delta$ a dimensionless quantity. By introducing the dimensionless time $\tau=M_p t$,  the following asymptotic 
behavior for the following dimensionless quantities is derived \cite{emelo0}
\be\lb{beja}
h=\frac{H}{M_p}=\frac{n}{\tau},\qquad a_0=\frac{A_0}{M_p^{\frac{3}{2}}}=k \tau^p,\qquad b_0=\frac{B_0}{M_p^{\frac{3}{2}}}=l \tau^p,
\ee
$$
q_a=\frac{Q_A}{M_p^{\frac{5}{2}}}=\sqrt{\dot{a}_0^2+3h^2 a_0^2}=k\tau^{p-1}\sqrt{p^2+3n^2},\qquad q_b=\frac{Q_B}{M_p^{\frac{5}{2}}}=\sqrt{\dot{b}_0^2+3h^2 b_0^2}=l\tau^{p-1}\sqrt{p^2+3n^2}.
$$
This behavior follows from a numerical analysis of equations (\ref{ecuac})-(\ref{ecuac2}) and (\ref{in}). Here $p\sim 3.6$ and $n\sim 2.1$ \cite{emelo0}. In terms of these quantities, the deviation from the Planck mass (\ref{real}) is given by
\be\lb{devi}
\delta M_p^2=M_p^2\bigg(\frac{a_0^2}{q_a}\frac{d\epsilon}{dq_a}+\frac{b_0^2}{q_b}\frac{d\epsilon}{dq_b}\bigg).
\ee
Now, for the solution (\ref{beja}) the function (\ref{alpe}) is asymptotically
$$
\epsilon=\frac{q_a^4-q_b^4}{q_a^2 q_b^2}+O(\delta).
$$
Thus, the mass deviation (\ref{devi}) is given by
\be\lb{devi2}
\delta M_p^2=2M_p^2\bigg(\frac{a_0^2}{q_a^2}-\frac{b_0^2}{q_b^2}\bigg)\frac{q_a^4+q_b^4}{q_a^2 q_b^2}.
\ee
Both terms in the parenthesis have a divergent behavior
 proportional to $\tau^2$.
However, it follows from (\ref{beja}) directly that
\be\lb{vang}
\frac{a_0^2}{q_a^2}=\frac{b_0^2}{q_b^2}\sim \frac{\tau^2}{p^2+3n^2},
\ee
and this, combined with (\ref{devi2}) implies that $\delta M_p^2\to 0$ for $\tau\to\infty$. In other words, this solution does not affect considerably the Newton constant $G_N$ at large times.

The argument given above has a problem. In the derivation of (\ref{real}) and (\ref{espin})-(\ref{ido2}) the time dependence of the coefficients of the perturbation lagrangian have been neglected. However, the coefficients
$a_0^2 r$ and $a_0^2s$ posses an asymptotic behavior
 of the form $\tau^2$, which is reflected in (\ref{vang}). So, the approximation just made may be non consistent.  However, we have checked that the resulting terms
that arise when the time dependence of these coefficients is not neglected are proportional to the time derivative of $\delta M_p^2$, which also tends to zero for large times. Thus, our conclusions are not modified for $t\to\infty$.

The next step is to analyze the gravitational wave issue for these models, which is a bit more complicated. This analysis requires the knowledge of the time dependence of the coefficients (\ref{busho}). In order to 
see this aspect, it is convenient to study the dependence of the quantities
$$
\frac{1}{q_a q_b}\frac{d^2\epsilon}{dq_b dq_a}=\frac{4q_b^4-4q_a^4}{q_b^4q_a^4},\qquad 
\frac{1}{q_a}\frac{d}{dq_a}\bigg(\frac{1}{q_a} \frac{d\epsilon}{dq_a}\bigg)=-\frac{8q_b^2}{q_a^6},
\qquad
\frac{1}{q_b}\frac{d}{dq_q}\bigg(\frac{1}{q_b}\frac{d\epsilon}{dq_b}\bigg)=\frac{8q_a^2}{q_b^6},
$$
with respect to the proper time at $t\to \infty$.
From here, by taking into account the asymptotic behavior
 (\ref{beja}), it follows that
$$
\frac{1}{q_a q_b}\frac{d^2\epsilon}{dq_b dq_a}\sim \frac{1}{\tau^{4p-4}},\qquad 
\frac{1}{q_a}\frac{d}{dq_a}\bigg(\frac{1}{q_a} \frac{d\epsilon}{dq_a}\bigg)\sim -\frac{1}{\tau^{4p-4}},
\qquad
\frac{1}{q_b}\frac{d}{dq_q}\bigg(\frac{1}{q_b}\frac{d\epsilon}{dq_b}\bigg)\sim \frac{1}{\tau^{4p-4}}.
$$
Therefore, the coefficients (\ref{busho}) have the following asymptotic behavior
\be\lb{aso}
\beta\sim b\sim\gamma\sim \frac{1}{\tau^{2p-2}}\frac{1}{\tau^{4n}},\qquad
\epsilon=\delta\sim e=d\sim\xi\sim\kappa\sim- \frac{1}{\tau^{2p-2}}\frac{1}{\tau^{2n}},
\qquad \lambda\sim \sigma\sim s\sim \frac{1}{\tau^{2p-2}}.
\ee
From here, by taking into account that asymptotically $A_0\sim B_0\sim \tau^p$, the following behavior of the coefficients
of the gravitational wave equations (\ref{inoa1})-(\ref{inoa5}) is inferred
$$
\bigg(\delta A_0+\frac{\kappa B_0}{2}\bigg)\sim\bigg(d B_0+\frac{\kappa A_0}{2}\bigg)\sim \frac{1}{\tau^{p-2}}\frac{1}{\tau^{2n}},\qquad
\bigg(\sigma A_0+\frac{\lambda B_0}{2}\bigg)\sim\bigg(s B_0+\frac{\lambda A_0}{2}\bigg)\sim \frac{1}{\tau^{p-2}},
$$
$$
\beta A_0 \sim \frac{1}{\tau^{p-2}}\frac{1}{\tau^{4n}},\qquad \frac{\gamma B_0}{2}\sim \frac{1}{\tau^{p-2}}\frac{1}{\tau^{4n}},\qquad (\epsilon A_0 + \frac{\xi B_0}{2})\sim \frac{1}{\tau^{p-2}}\frac{1}{\tau^{2n}},
$$
$$
b B_0\sim\frac{1}{\tau^{p-2}}\frac{1}{\tau^{4n}},
\qquad \frac{\gamma A_0}{2}\sim\frac{1}{\tau^{p-2}}\frac{1}{\tau^{4n}}, \qquad (e B_0+\frac{\xi A_0}{2})\sim \frac{1}{\tau^{p-2}}\frac{1}{\tau^{2n}},
$$ 
$$
\epsilon A_0\sim \frac{1}{\tau^{p-2}}\frac{1}{\tau^{2n}},\qquad \frac{\epsilon A_0^2}{4}\sim \frac{1}{\tau^{2n-2}},\qquad\frac{\delta A_0^2}{4}\sim \frac{1}{\tau^{2n-2}},\qquad 
eB_0\sim \frac{1}{\tau^{p-2}}\frac{1}{\tau^{2n}},\qquad \frac{e B_0^2}{4}\sim\frac{1}{\tau^{2n-2}},
$$
$$
\frac{d B_0^2}{4}\sim\frac{1}{\tau^{2n-2}},
\qquad \frac{\xi B_0}{2} \sim \frac{1}{\tau^{p-2}}\frac{1}{\tau^{2n}},\qquad \frac{\xi A_0}{2}\sim  \frac{1}{\tau^{p-2}}\frac{1}{\tau^{2n}},\qquad  \frac{\xi A_0B_0}{4}\sim\frac{1}{\tau^{2n-2}},
$$
\be\lb{frage}
 \frac{\kappa A_0B_0}{4}\sim\frac{1}{\tau^{2n-2}},\qquad \frac{\gamma A_0 B_0}{2}\sim \frac{1}{\tau^{4n-2}},\qquad
\frac{\beta A_0^2}{2} \sim \frac{1}{\tau^{4n-2}},
\qquad
\frac{b B_0^2}{2}\sim\frac{1}{\tau^{4n-2}},
\ee
$$
\frac{\lambda A_0B_0}{2}\sim \tau^2,\qquad
\frac{\sigma A_0^2}{2}\sim \tau^2,\qquad \frac{s B_0^2}{2}\sim \tau^2.
$$
We have not included the coefficients $\alpha$ and $a$ given by
\be\lb{aaa}
\alpha=\frac{1}{2Q_A}\frac{d\epsilon}{dQ_A},\qquad a=\frac{1}{2Q_B}\frac{d\epsilon}{dQ_B},
\ee
since we have seen above that they do not go to zero for $\tau\to\infty$. The same consideration applies for the last three coefficients in (\ref{frage}), which go like $\tau^2$.
By neglecting the other terms, it follows that (\ref{inoa2}) become asymptotically
$$
2 (\alpha c_{i;\beta})^{;\beta}-\bigg(\frac{\alpha A_0}{2}h_{00;i}\bigg)_{;0}+\frac{\alpha A_0}{2}h_{ij;0j}\simeq 0,
$$
\be\lb{inua2}
2 (a d_{i;\beta})^{;\beta}-\bigg(\frac{a B_0}{2}h_{00;i}\bigg)_{;0}+\frac{a B_0}{2}h_{ij;0j}\simeq 0.
\ee
From the two equations (\ref{inua2}) it is inferred that
\be\lb{reo}
\frac{c_i}{A_0}\simeq \frac{d_i}{B_0}=f(h_{ij}).
\ee
This relation is completely schematic, but it states that the behavior of $c_i$ and $d_i$ are determined by the same quantity $f(h_{ij})$, which depends on the metric components $h_{ij}$, up to a proportionality
factor $A_0$ or $B_0$.  Now, by use of this and the formulas (\ref{frage}) it follows that in both equations (\ref{inoa1}) the dominant term for large $\tau$ is the one proportional to $\alpha A_0$ or $aA_0$.
Thus, both equations give asymptotically that $\bigg(\alpha A_0 h_{00;\alpha}\bigg)^{;\alpha}=0$ at very large times. However, it will be useful to consider very large but finite times.  By the first formulas (\ref{frage})  it follows that the next order dominant term 
is proportional to $2\sigma A_0+\lambda B_0$ and $2s B_0+\lambda A_0$, together with $\sigma c_{0;0}$
and $\lambda c_{0;0}$. In this approximation the equations (\ref{inoa1}) reduce to
$$
\bigg[\bigg(\sigma A_0+\frac{\lambda B_0}{2}\bigg)h_{00;0}+2\sigma c_{0;0}\bigg]_{;0}\simeq-\bigg(\frac{\alpha A_0}{2}h_{00;\alpha}\bigg)^{;\alpha},
$$
\be\lb{inua1}
\bigg[\bigg(s B_0+\frac{\lambda A_0}{2}\bigg)h_{00;0}+2\lambda c_{0;0}\bigg]_{;0}\simeq-\bigg(\frac{a B_0}{2}h_{00;\alpha}\bigg)^{\alpha}.
\ee
On the other hand the asymptotic form of the  equation (\ref{inoa3}) is 
$$
-(\alpha A_0 c^{(0;\alpha)})_{;\alpha}
+(\alpha A_0^2 h^{00;\alpha})_{;\alpha}-(\frac{\alpha A_0^2}{2}h^{00;0})_{;0}
-(a B_0 d^{(0;\alpha)})_{;\alpha}
+(a B_0^2 h^{00;\alpha})_{;\alpha}-(\frac{a B_0^2}{2}h^{00;0})_{;0}
$$
\be\lb{inua3}
+\frac{1}{2}\bigg[\bigg(\sigma A^2_0+\frac{\lambda B_0A_0}{2}\bigg)h_{00;0}+2\sigma A_0 c_{0;0}\bigg]_{;0}+\frac{1}{2}\bigg[\bigg(s B_0^2+\frac{\lambda A_0 B_0}{2}\bigg)h_{00;0}+2\lambda c_{0;0}B_0\bigg]_{;0}+E_{00}\simeq 0.
\ee
Here the term $E_{00}$ is the contribution of $\delta L_{EH}$. Now, it will be shown that several of the terms in (\ref{inua3}) cancel asymptotically. In order
to visualize this, note that the first and the fourth term of the equation (\ref{inua3}) combine as
\be\lb{reo2}
-(\alpha A_0 c^{(0;\alpha)})_{;\alpha}-(a B_0 d^{(0;\alpha)})_{;\alpha}\simeq-[(\alpha A_0^2+a B_0^2) f(h_{ij})^{;\alpha}]_{;\alpha},
 \ee
 where in the last line the relation (\ref{reo}) was taken into account. By bearing in mind the formulas (\ref{aaa}) it follows that the quantity in parenthesis in (\ref{reo2})
 is proportional to the quantity given in (\ref{devi}), that is 
 \be\lb{canso}
 \alpha A_0^2+a B_0^2 \sim\frac{a_0^2}{q_a}\frac{d\epsilon}{dq_a}+\frac{b_0^2}{q_b}\frac{d\epsilon}{dq_b}.
 \ee
  But the right hand side has been shown in (\ref{devi}) to vanish for large times $t\to\infty$, since it is proportional to the deviation of the Newton constant $\delta M_p^2$. Thus the contribution for this pair of terms for asymptotic times is negligible. The same consideration follows for second and the fifth term of (\ref{inua3}), and the third and the sixth. 
It remains to see that the seventh and the eight terms of (\ref{inua3}) cancel each other. To analyze this, consider again the equations (\ref{inua1}).
The first of these equations may be rewritten as
$$
\bigg[\bigg(\sigma A_0+\frac{\lambda B_0}{2}\bigg)h_{00;0}+2\sigma c_{0;0}\bigg]_{;0}\simeq \bigg(\frac{\alpha A_0}{2}h_{00;0}\bigg)^{;0}-\frac{\alpha A_0}{2} h_{00;ii}.
$$
Equipped with this equation, one would like to analyze the seventh term of (\ref{inua3}). This term is elaborated, by use of the last identity, as follows
$$
\bigg[\bigg(\sigma A^2_0+\frac{\lambda B_0A_0}{2}\bigg)h_{00;0}+2\sigma A_0 c_{0;0}\bigg]_{;0}\simeq -A_0\bigg[\bigg(\frac{\alpha A_0}{2}h_{00;0}\bigg)^{;0}-\frac{\alpha A_0}{2} h_{00;ii}\bigg]
$$
$$
-\frac{pA_0}{\tau}\bigg[\frac{\alpha A_0}{2}h_{00;0}-\int \frac{\alpha A_0}{2} h_{00;ii}dt\bigg]=-A_0\bigg(\frac{\alpha A_0}{2}h_{00;0}\bigg)^{;0}+\frac{\alpha A^2_0}{2} h_{00;ii}
$$
$$
-\frac{pA_0}{\tau}\bigg[\frac{\alpha A_0}{2}h_{00;0}-\frac{\alpha \tau A_0}{2p} h_{00;ii}+\frac{\alpha \tau A_0}{2p} \int h_{00;ii0}dt\bigg].
$$
Here, the fact that $A_0\sim \tau^p$ has been taken into account. By using the integration part formula on the first term of the last equality, one obtains finally
$$
\bigg[\bigg(\sigma A^2_0+\frac{\lambda B_0A_0}{2}\bigg)h_{00;0}+2\sigma A_0 c_{0;0}\bigg]_{;0}\simeq-\bigg(\frac{\alpha A^2_0}{2}h_{00;0}\bigg)^{;0}+\frac{\alpha pA^2_0}{2\tau}h_{00;0}+\frac{\alpha A^2_0}{2} h_{00;ii}
$$
\be\lb{jamon1}
-\frac{\alpha pA^2_0}{2\tau}h_{00;0}+\frac{\alpha A^2_0}{2} h_{00;ii}-\frac{\alpha A^2_0}{2} \int h_{00;ii0}dt.
\ee
In the same fashion, from the second (\ref{inua1}) it is obtained that the eight term of (\ref{inua3}) is given by
$$
\bigg[\bigg(s B^2_0+\frac{\lambda B_0A_0}{2}\bigg)h_{00;0}+2\lambda B_0 c_{0;0}\bigg]_{;0}\simeq-\bigg(\frac{a B^2_0}{2}h_{00;0}\bigg)^{;0}+\frac{a pB^2_0}{2\tau}h_{00;0}+\frac{a B^2_0}{2} h_{00;ii}
$$
\be\lb{jamon2}
-\frac{a pB^2_0}{2\tau}h_{00;0}+\frac{a B^2_0}{2} h_{00;ii}-\frac{a B^2_0}{2} \int h_{00;ii0}dt.
\ee
Clearly both (\ref{jamon1})-(\ref{jamon2}) have the same structure, except for a proportionality factor $\alpha A_0^2$ and $aB_0^2$. In other words
$$
\bigg[\bigg(s B^2_0+\frac{\lambda B_0A_0}{2}\bigg)h_{00;0}+2\sigma A_0 c_{0;0}\bigg]_{;0}+\bigg[\bigg(\sigma A^2_0+\frac{\lambda B_0A_0}{2}\bigg)h_{00;0}+2\lambda B_0 c_{0;0}\bigg]_{;0}\sim \alpha A_0^2+a B_0^2 \to 0,
$$
for $\tau\to \infty$.  Thus, equation (\ref{inua3}) reduces to $E_{00}=0$, which is one of the standard equation of gravitational waves in GR.

In addition, for the component $h_{0k}$ the equation (\ref{inoa4}) reduces to
$$
- (\alpha A_0c^{(k;\alpha)})_{;\alpha}+(\alpha A_0c^{(0;\alpha)})_{;0}
+\bigg(\frac{\alpha A_0^2}{2}h^{0k; \alpha}\bigg)_{;\alpha}+\bigg(\frac{\alpha A_0^2}{2}h^{0\alpha;k}\bigg)_{;\alpha}-\bigg(\frac{\alpha A_0^2}{2}h^{k\alpha;0}\bigg)_{;\alpha}-\bigg(\frac{\alpha A_0^2}{2}h^{00;k}\bigg)_{;0}
$$
$$
- (aB_0d^{(k;\alpha)})_{;\alpha}+(a B_0d^{(0;\alpha)})_{;0}
+\bigg(\frac{a B_0^2}{2}h^{0k; \alpha}\bigg)_{;\alpha}+\bigg(\frac{a B_0^2}{2}h^{0\alpha;k}\bigg)_{;\alpha}-\bigg(\frac{a B_0^2}{2}h^{k\alpha;0}\bigg)_{;\alpha}-\bigg(\frac{a B_0^2}{2}h^{00;k}\bigg)_{;0}+E_{0k}=0.
$$
By use of (\ref{canso}) and (\ref{reo}) it is seen again that all the terms cancel when chosen in pairs conveniently, and the equation reduces to $E_{0k}=0$, which is another standard gravitational wave equation
in GR.  Finally, for $h_{ik}$, it is obtained that
$$
(\alpha A_0c^{(k;l)})_{;0}+\bigg(\frac{\alpha A_0^2}{2}h^{lk;0}\bigg)_{;0}
+(aB_0d^{(k;l)})_{;0}+\bigg(\frac{a B_0^2}{2}h^{lk;0}\bigg)_{;0}+E_{kl}=0.
$$
Again, all the terms cancel and one obtains asymptotically that $E_{kl}=0$. Thus, all the standard gravitational waves equations of GR have been obtained, under the approximations made.
Based on this, we suggest that there are no large deviations from transversality or from the standard dispersion relation $\omega=k$ of the gravitational waves for asymptotic times $t\to \infty$ for the solutions presented in \cite{emelo0}.

\section{New solutions}
In the previous section, it was shown that for a generic behavior
 (\ref{beja}) with $p>2$ and $n>2$, the resulting dynamics will not generate neither an unacceptable effective gravitational constant $G_N$ nor 
longitudinal gravitational waves for large times.  Thus, there is no reason for restricting  the attention to the lagrangian density (\ref{alpe}). One can consider any other lagrangian density $\epsilon(Q_A, Q_B)$ which give rise to any behavior
 (\ref{beja}) with $p>2$ and $n>2$. For instance, consider the following lagrangian density
\be\lb{novoco}
\epsilon(Q_A, Q_B)=\frac{Q_A^{2m}-Q_B^{2m}}{\delta M_p^{2m}+Q_A^m Q^m_B}.
\ee
Here $m$ is, at this point, an arbitrary power.
By assuming that the solutions corresponding to this model posses the asymptotic behavior
 (\ref{beja}), then by  plugging this behavior
 it into the equations of motion (\ref{ecuac})-(\ref{ecuac2}) and (\ref{in})
one should obtain a relation between the exponents in (\ref{beja}). The dimensionless version of the equations of motion  (\ref{ecuac})-(\ref{ecuac2}) and (\ref{in}) are the following
\be\lb{uan}
\bigg[(\ddot{a}_0+ 3\,h\,\dot{a}_0-3\, h^2\, a_0
)\,\frac{de}{q_A\,d q_A}
+\dot{a}_0\,\frac{d}{d\tau} \bigg(\frac{de}{q_A\,d q_A}\bigg)
\bigg]=0,
\ee
\be\lb{tu}
\bigg[\;\Big( \ddot{b}_0+ 3\,h\,\dot{b}_0-3\, h^2\, b_0
\Big)\,\frac{de}{q_B\,d q_B}
+\dot{b}_0\,\frac{d}{d\tau} \Big(\frac{de}{q_B\,d q_B}\Big)
\bigg]=0,
\ee
$$
2\,\dot{h} + 3\, h^2 =
\lambda+\bigg[\;\widetilde{e}(q_A,\,q_B)
-\frac{d}{d\tau}\Big(h\,a_0^2\,\frac{de}{q_A\,d q_A})
+ \dot{a}_0^2\,\frac{de}{q_A\,d q_A}
$$
\be\lb{foor}
-\frac{d}{d\tau}\Big(h\,b_0^2\,\frac{de}{q_B\,d q_B})
+ \dot{b}_0^2\,\frac{de}{q_B\,d q_B}\;
\bigg],
\ee
\be\lb{zri}
3\, h^2 =\lambda+\widetilde{e}(q_A, q_B),\qquad  q_A=\sqrt{\dot{a}_0^2+3\,h^2\, a_0^2},\qquad q_B=\sqrt{\dot{b}_0^2+3h^2 b_0^2}.
\ee
Here the overdot stands for differentiation with respect to $\tau$.
When the asymptotic solution (\ref{beja}) is considered, one has that
$$
\frac{1}{q_A} \frac{\mathrm{de} }{\mathrm{d} q_A} = \frac {n} {\tau^{2(p-1)}} \frac {(A^{m-2}B^{3m}+A^{3m-2}B^{m})} {A^{2m}B^{2m}}+2\delta \frac {m}{\tau^{(2m+2)(p-1)}} \frac {A^{2m-2}}{A^{2m}B^{2m}},
$$
\be\lb{agus}
\frac{1}{q_B} \frac{\mathrm{de} }{\mathrm{d} q_B} =- \frac {n} {\tau^{2(p-1)}} \frac {(B^{3m-2}A^{m}+A^{3m}B^{m-2})} {B^{2m}A^{2m}}-2\delta \frac {m}{\tau^{(2m+2)(p-1)}} \frac {B^{2m-2}}{A^{2m}B^{2m}}.
\ee
Here, the following coefficients have been introduced
\be\lb{cofa}
A=k\sqrt{p^2+3n^2},\qquad
B=l\sqrt{p^2+3n^2}.
\ee
By taking into account (\ref{agus}), it follows after some algebraic operations that the asymptotic behavior
 (\ref{beja}) is compatible with equations (\ref{uan})-(\ref{tu}) when the following algebraic relation is satisfied 
\be\label{constantes}
p(p-1)-3np+3n^2=0.
\ee
On the other hand, the quantity $\widetilde{e}(q_A, q_B)$ introduced in (\ref{zri}) can be worked out as
$$
\widetilde{e}(q_A, q_B)=e-q_A \frac{de}{d q_A}-q_B \frac{de }{d q_B}= \frac {[q_A^mq_B^m+\delta(1-2m)][q_A^{2m}-q_B^{2m}]}{[\delta +q_A^mq_B^m]^2}.
$$
By taking the last equation into account, it follows that the Friedmann equation (\ref{zri})  becomes
$$
\frac{3n^2}{\tau^2}= \lambda+\frac{q_A^{2m}-q_B^{2m}}{q_A^mq_B^m}.
$$
By requiring that this equation holds when $\tau\to\infty$, it is obtained that
\be
\lambda+\frac{A^{2m}-B^{2m}}{A^mB^m}= \lambda+\frac{k^{2m}-l^{2m}}{k^ml^m}=0,
\ee
where in the last step the definition (\ref{cofa}) has been taken into account.
The solution of this algebraic equation gives
\be\label{constantekl}
\frac {k^m}{l^m}= \sqrt{(\frac{\lambda}{2})^2+1}-\frac{\lambda}{2}.
\ee
The final equation to solve is (\ref{foor}). For working out this equation one should calculate first that
$$
\frac{d }{d \tau}(ha_0^2\frac{1}{q_a} \frac{de }{dq_A})=nk^2n\frac{A^{m-2}B^{3m}+A^{3m-2}B^{m}}{A^{2m}B^{2m}}+\frac{2\delta nmk^2}{\tau^{{2m(p-1)}}}\frac{A^{2m-2}}{A^{2m}B^{2m}} [2m(1-p)+1],
$$
$$
\frac{d }{d\tau}(h b_0^2\frac{1}{q_b} \frac{de }{d q_B})=
-nl^2n\frac{A^{m}B^{3m-2}+A^{3m}B^{m-2}}{A^{2m}B^{2m}}-\frac{2\delta nml^2}{\tau^{{2m(p-1)}}}\frac{B^{2m-2}}{A^{2m}B^{2m}} [2m(1-p)+1].
$$
In these terms, equation (\ref{foor}) simplifies by noticing that $\lambda+\widetilde{e}$ vanishes at zero order in  $\delta$, so only the part of $\widetilde{e}$ to be considered is the first order one.
After some elementary calculations, it is found that
the terms of this equations which do not contain $\delta$ vanish identically, and therefore 
$$
 \frac {(3n-2)n}{\tau^2}
=\frac{\delta(p^2+3n^2)^{m-1}(k^{2m}-l^{2m})}{A^{2m}B^{2m}\tau^{2m(p-1)}}   [2p^2m+2nm (2mp-2m+1)+(1-2m)(p^2+3n^2)].
$$
Since the values of $n$ and $p$ are not known, then one should insure that the right hand of the equation goes to $0$ when  $\tau\rightarrow\infty$. 
This can be shown to be, after some algebraic manipulations, the following condition
\be\lb{mor2}
p^2 + nm [4m (p-1)-2]+3n^2(1-2m)=0.
\ee
This algebraic equation insures that the right hand is zero. Both equations (\ref{constantes}) and (\ref{mor2}) are equivalent to the system
$$
(4m^2n+3n+1)p=6n^2m+4n m^2+2nm,
$$
$$
(6n^2m+4n m^2+2nm)p=2m(1+2m+6n+6nm+9n^2)+3n+9n^2+12m^2n^2.
$$
This system has the compatibility condition
$$
a_3 n^3+a_2 n^2+a_1 n+a_0=0,
$$
with coefficients
$$
a_0=-2m(1+2m),\qquad
a_1=-4m^2(1+2m)^2-3(1+4m+4m^2)-2m(1+2m)(3+4m^2),
$$
$$
a_2=24m^2(1+2m)-3(3+4m^2)(1+4m+4m^2)-3(3+6m+4m^2),
$$
$$
a_3=36m^2-3(3+4m^2)(3+6m+4m^2).
$$
For instance, for $m=0.89$, it is calculated that $n=2.5$ and $p=2.6$. This solution pass the test of the previous section since $p>2$ and $n>2$.
There are several other solutions with this behavior. Thus, we have enlarged the list of possible adjustment mechanisms initiated in \cite{emelo0}. However, the best values for $n$ from the phenomenological point of view  are in the interval $[0.5, 0.66]$. In view of this, we have to emphasize that the condition $p>2$ and $n>2$ is sufficient, but not necessary. We have considered this case in order to simplify the analysis. 
For other situations, there may be a cancellation but the analysis is much more involved. We leave this for a future investigation.

\section{Discussion and open questions}

In the present work some evidence was collected that suggest that the vector adjustment vector fields presented in \cite{emelo0}
do not affect the Newtonian dynamics of the planets at present times. Furthermore, it was suggested her that the gravitational
waves that these models predict are asymptotically transversal and with the right dispersion relation $\omega=k$.
In the process for checking this some new models were found. The models interpolate between an initial de Sitter universe and a Friedmann-Robertson-Walker type.  
It would be be interesting to study
these matters for the models \cite{emelo1}-\cite{emelo4}, which seem to be more general than these ones. Another open question is 
to interpret the lagrangian $\epsilon(Q_A, Q_B)$ presented as an effective one, and to identify an underlying theory. Of course this may not be a simple task, but perhaps
the Bjorken ideas \cite{bjorken1}-\cite{bjorken2} may be useful in this context.

There exist several open problems to be investigated further. The present paper address the possibility of making the effective cosmological small, but
it does not explain why the actual energy density  is so close to the critical one $\rho_c$. The value of $\rho_c$ has an uncanny relation within
the pion mass $m_\pi$ and the Planck mass $M_p$. There exist some speculations that interpret this as a result of an interaction between the hadronic sector
and a hidden one, which can be very weak, of gravitational order \cite{calo}-\cite{gaba}. It may be interesting to find a combined model with includes an adjustment field 
which cancel the large contribution of the different QFT to the universe energy density and leave the contribution of this extremely weak interaction untouched at large times.
We leave this for a future investigation.

\section*{Acknowledgments}
O. P. S is supported by the CONICET.

\end{document}